\documentclass[]{amsart}
\usepackage{times}
\usepackage[authoryear]{natbib}
\usepackage[]{graphicx}

\makeatletter
\def\paragraph{\@startsection{paragraph}{4}%
  \z@\z@{-\fontdimen2\font}%
  {\normalfont\bfseries}}
\makeatother

\title{Robust incorporation of historical information with known type I error rate inflation}
\author{Silvia Calderazzo} 
\author{Annette Kopp-Schneider}
\date{\today}

\begin{document}

\maketitle

\begin{abstract}
Bayesian clinical trials can benefit of available historical information through the elicitation of informative prior distributions. Concerns are however often raised about the potential for prior-data conflict and the impact of Bayes test decisions on frequentist operating characteristics, with particular attention being assigned to inflation of type I error rates. 
This motivates the development of principled borrowing mechanisms, that strike a balance between frequentist and Bayesian decisions. Ideally, the trust assigned to historical information defines the degree of robustness to prior-data conflict one is willing to sacrifice. However, such relationship is often not directly available when explicitly considering inflation of type I error rates. 
We build on available literature relating frequentist and Bayesian test decisions, and investigate a rationale for inflation of type I error rate which
explicitly and linearly relates the amount of borrowing and the amount of type I error rate inflation in one-arm studies.
A novel dynamic borrowing mechanism tailored to hypothesis testing is additionally proposed. We show that, while dynamic borrowing prevents the possibility to obtain a simple closed form type I error rate computation, an explicit upper bound can still be enforced. Connections with the robust mixture prior approach, particularly in relation to the choice of the mixture weight and robust component, are made. Simulations are performed to show the properties of the approach for normal and binomial outcomes.
\end{abstract}

\section{Introduction}

While the adoption of Bayesian clinical trial designs is on the rise, particularly in early phase trials, additional reporting and `reasonable' control of frequentist operating characteristics has often been a requirement of regulatory authorities  \citep[][]{fda}. A key advantage of Bayesian designs is the possibility to include in the analysis relevant historical information about the model parameters through the elicitation of informative prior distributions. If such information is consistent with the information collected during the current trial, improvement of the design operating characteristics, e.g. in terms of test error rates and estimation error, can generally be achieved. However, if prior-data conflict is present, losses can be significant. To deal with this problem, robustification approaches have been proposed, in the form of static or dynamic (i.e. depending on the currently observed data) discounting of the historical information conveyed by the prior. While both static and dynamic robustification approaches can decrease the impact of prior-data conflict, no power gains can be achieved when strict control of type I rate is required if a uniformly most powerful (UMP) test  is available \citep{kopp2019}. 
Available robustification approaches, e.g., the robust mixture prior \citep[see e.g.][]{berg1986,schm2014}, the power prior  \citep[see e.g.][]{ibra2000}, the commensurate prior \citep[see e.g.][]{hobb2012} require the specification of additional parameters and/or distributions. Such choices, while fully characterised from a Bayesian prospective where prior beliefs have their own right, are not typically intuitively related to their impact on the control of frequentist operating characteristics. Selection of a borrowing weight based on explicit control of type I error rate, is investigated in the power prior context by \cite{niko2017}. The approach does however not allow a simple interpretation of the weight in terms of type I error rate inflation.

The purpose of the present work is to investigate and characterise a simple and easily interpretable compromise solution between Bayesian and frequentist test decisions controlling type I error rate at a pre-specified `standard' level in one-arm studies. The proposed compromise solution can be obtained as a Bayes test decision with adjusted posterior probability test thresholds, or equivalently as a frequentist test decision with an adjusted rejection region. The solution is derived in the spirit of \cite{berg1985} who states that in a normal prior - normal likelihood case \textit{`the Bayesian method can be  thought of as providing a rational way of choosing the size of the test'}, but with the aim of achieving a gradual compromise between Bayesian and frequentist test decisions. A compromise of this type was first investigated in \citet{hodg1952}, where restricted Bayes solutions were introduced as a general tool to characterise decisions minimising the integrated risk under a maximum frequentist risk constraint. While the core principle, i.e., a compromise on test decision thresholds, stays the same, we construct such a compromise by introducing a weighting of historical information which linearly relates to the type I error rate inflation. 
The proposed approach can also be used as a dynamic borrowing mechanism, and in this context we propose a novel approach to adaptively select the weight assigned to external information according to the agreement between historical and current information in supporting the alternative hypothesis. 

Section \ref{sec:dectheory} outlines the decision-theoretic background and reviews the restricted Bayes approach of \citet{hodg1952}. Section \ref{sec:proposedAppr} describes the proposed approach and its relationship to other robust borrowing approaches. Section \ref{sec:simexample} presents exemplary applications to simulated data for normal outcomes. We conclude the article with a discussion of the results and an outlook on future work.

\section{Hypothesis testing under the weighted 0-1 loss}
\label{sec:dectheory}

\subsection{Definitions and set-up}
\label{sec:hptest}

We shortly introduce the core concepts and notation which will be used in the following. Note that a similar setup has also been considered in \cite{cald2020} to which we refer for additional details.
We consider the situation of testing $H_0:\theta \leq \theta_0$ versus the alternative hypothesis $H_1:\theta> \theta_0$. We denote by $\mathbf{y}$ the observed data having probability density function $f(\mathbf{y}|n,\theta)$ indexed by $\theta$ and sample size $n$, and by $\pi(\theta)$ the prior distribution for $\theta$. 
A test decision $d$ has to be taken for rejecting ($d=1$) or keeping ($d=0$) the null hypothesis. A 0-$\kappa$ \emph{loss function} $L(\theta, d_{\kappa}(\mathbf{y}))$ assigns unit cost to a type I error, a cost $\kappa$ to a type II error, and no cost if no test error is made. Note that only the ratio between the cost of a type I and a type II error is relevant, so the unit cost for the type I error is assumed without loss of generality.

If no prior information or belief about $\theta$ is available, a test decision can be undertaken following the frequentist approach. Here, the aim is to minimise over the whole parameter space $\Theta$ the maximum of the \emph{frequentist risk} 
\begin{equation*}
R(\theta, d_{\kappa}(\mathbf{y})) = I(\theta \leq \theta_0) \beta(\theta) + \kappa I(\theta >\theta_0) (1- \beta(\theta)), 
\end{equation*}
where
\begin{equation*}
\beta(\theta)=P^f(d_{\kappa}(\mathbf{y})=1|\theta)
\end{equation*}
denotes the rejection probability, $I$ is the indicator function, and $P^f$ represents the probability computed with respect to the distribution $f$, i.e., the frequentist risk is obtained by averaging the loss function with respect to the data density $f$. The maximum of $R(\theta, d_{\kappa}(\mathbf{y}))$ over $\Theta$ is given by the maximum between $ \beta(\theta_0)$ and $ \kappa \{1-\beta(\theta_0)\}$, if $\beta(\theta)$ is strictly increasing in $\theta$, as in, e.g., the case of normally distributed observations, see Figure 1. The maximum frequentist risk is then minimised by taking $d_{\kappa}(\mathbf{y})$ such that $ \beta(\theta_0)=\kappa \{1- \beta(\theta_0)\}$. Such a decision will be referred to in the following as `Frequentist decision' (FD). It follows that the maximum type I error rate $\beta(\theta_0)$ is equal to $\tau=\kappa/(1+\kappa)$. The frequentist approach is often described as an `ultra-pessimistic' approach, which is effectively aimed at minimising the loss under the worst possible scenario.

The Bayesian approach associates uncertainty to the parameter while conditioning on the observed data, and thus aims at minimising the \emph{posterior expected loss} 
\begin{equation*}
\rho(\pi,  d_{\kappa}(\mathbf{y})| \mathbf{y})= P^{\pi}(\theta \leq \theta_0|\mathbf{y}) I(d_{\kappa}(\mathbf{y})=1) + \kappa P^{\pi}(\theta > \theta_0|\mathbf{y}) I(d_{\kappa}(\mathbf{y})=0).
\end{equation*}
 The posterior expected loss $\rho$ is obtained by averaging the loss function with respect to the posterior density $\pi(\theta|\mathbf{y})$, and
is minimised by the `Bayes decision' (BD) with respect to prior $\pi$, $ d_{\kappa}^{\pi}(\mathbf{y})$, which rejects $H_0$ if 
\begin{equation*}
P^{\pi}(\theta \leq \theta_0|\mathbf{y})<\tau
\end{equation*}
with $\tau= \kappa/(1+\kappa)$. The underlying assumption is that $\pi$ is a reliable summary of the available information (or prior beliefs) about the parameter $\theta$.
Uncertainty about the data outcomes can be additionally taken into account by computing the \emph{integrated risk} 
\begin{equation*}
\nonumber r(\pi, d_{\kappa}(\mathbf{y}))=\int_{-\infty}^{\theta_0} \beta(\theta)  \pi(\theta) d\theta + \kappa \int_{\theta_0}^{\infty}  (1-\beta(\theta)) \pi(\theta) d\theta,
\end{equation*}
where the loss function is averaged with respect to both the data and the prior distribution.
The integrated risk is also minimised by the BD $d_{\kappa}^{\pi}(\mathbf{y})$ which minimises the posterior expected loss for each $\mathbf{y} \in Y$ \citep[see e.g.][]{robe2007}, and leads to a rejection probability $\beta^{\pi}(\theta)=P^f(d^{\pi}_{\kappa}(\mathbf{y})=1|\theta)$.

\begin{figure}[h!]
   \centering
   \includegraphics[width=1\textwidth]{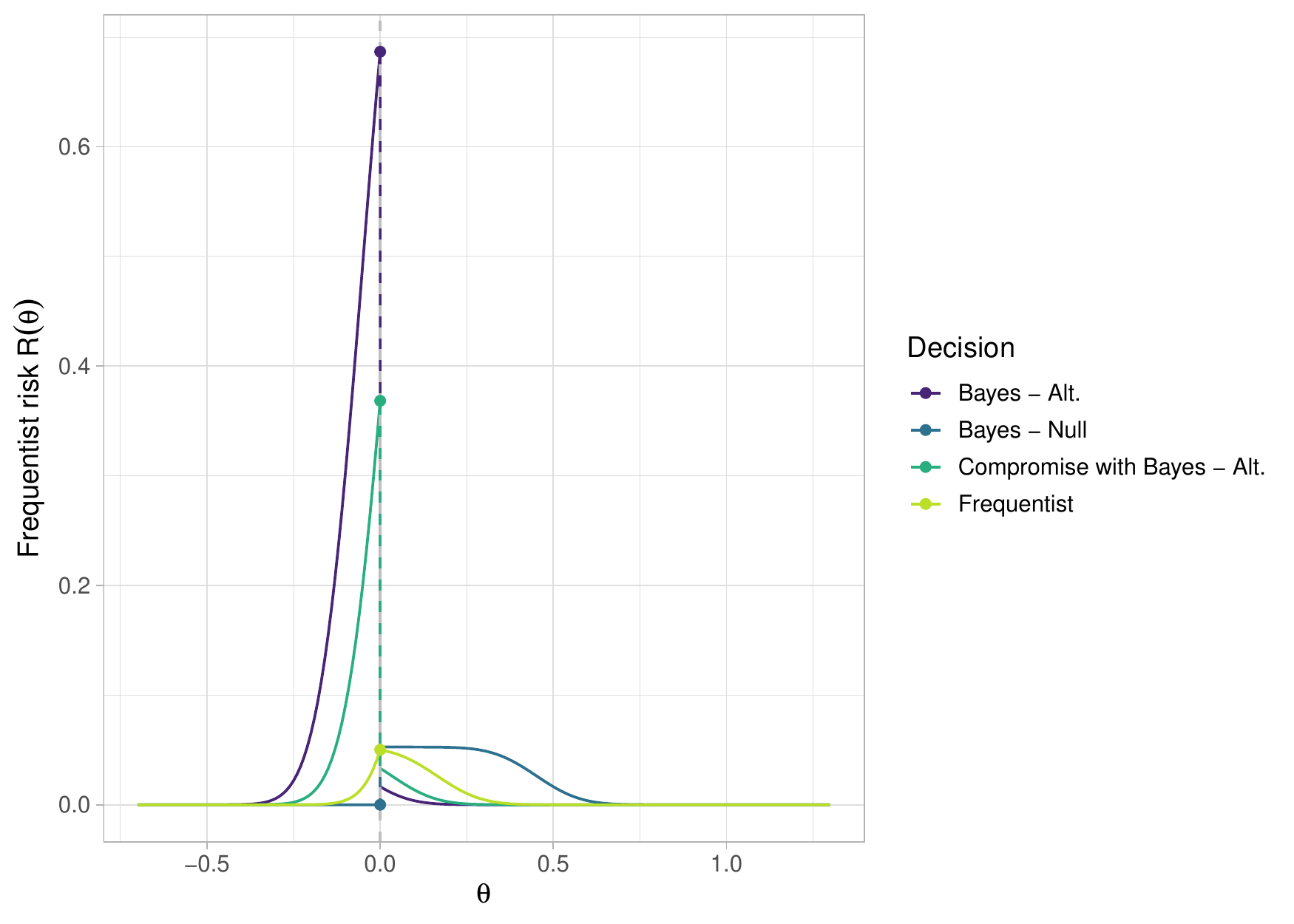}
   \caption{Frequentist risk $R(\theta, d_{\kappa})$ under the 0-$k$ loss for a normal endpoint mean with unit standard deviation and $n=100$, $\kappa=0.05/0.95$, $\theta_0=0$. $R(\theta, d_{\kappa})$ is computed under the FD, the BD under prior $N(-0.5,\sqrt{1/50})$ (`Bayes-Null'), the BD under prior $N(0.5,\sqrt{1/50})$ (`Bayes-Alt.'), and the CD between the FD and the BD `Bayes-Alt.' supporting $H_1$ with weight $w=0.5$.}
\label{fig:risk_ex}
\end{figure}

\subsection{Restricted Bayes decisions}
\label{sec:restrbayes}

Often, a compromise between frequentist decisions and Bayes decisions induced by an informative prior distribution is sought. In the proposal of \citet{hodg1952} this compromise is characterised as a `restricted Bayes solution' (henceforth referred to as the `restricted Bayes decision', RBD), in which the optimal decision is the one minimising the integrated risk with respect to an (informative) prior distribution $\pi$, subjected to a constraint on the maximum frequentist risk. Formally, $d^{\nu}$ is defined to be a restricted Bayes decision with respect to a prior distribution $\pi$  subjected to $\underset{\theta \in \Theta}{\sup} R(\theta, d) = C$ if, for a given constant $\eta \in [0,1]$, it minimises 
\begin{equation}
    \eta r(\pi, d) + (1-\eta) \ \underset{\theta \in \Theta}{\sup} \ R(\theta, d), \text{ and} \ \underset{\theta \in \Theta}{\sup} R(\theta, d^{\nu}) = C.
    \label{eq:restrBayes}
\end{equation}
The value of $C$ thus represents the maximum frequentist risk one is willing to accept when incorporating prior information in the decision.
The restricted Bayes decision can also be identified as the Bayes decision with respect to a specific prior distribution, i.e. a distribution of the form $\nu= \eta \pi + (1-\eta) \mu_0$, for given constant $\eta \in [0,1]$ and distribution $\mu_0$, such that $\int_\Theta R(\theta, d^\nu) \ \mu_0(\theta) \ d\theta = \underset{\theta \in \Theta}{\sup} R(\theta, d^\nu)$ \citep{hodg1952}. In many problems, the identification of $\mu_0$ and $\eta$ is nontrivial, and thus approximate restricted Bayes decisions have been sought \citep[see][]{berg1985}.  

Since we are here in the context of hypothesis testing, identification of $\mu_0$ is more straightforward. If the frequentist risk has a unique maximum, then necessarily $\mu_0$ corresponds to a point-mass prior assigning all probability to the point at which such maximum is achieved \citep{bayr2011}. However, in the context of clinical trial design, interest is typically placed on constraining type I error rate, rather than the maximum frequentist risk.
If the maximum type I error rate is achieved at $\theta_0$ (i.e. $\beta(\theta)$ is strictly increasing in $\theta$), then we define the `type I error rate restricted Bayes decision' (TI-RBD) as the one minimising
\begin{equation}
    \eta r(\pi, d_{\kappa}) + (1-\eta) R(\theta_0, d_{\kappa}),
    \label{eq:restrictedBayes}
\end{equation}
for a given $\eta \in [0,1]$, and corresponding to the Bayes decision with respect to $\nu= \eta \pi + (1-\eta) \mu_0$, where $\mu_0$ is the point-mass distribution at ${\theta_0}$ (see Appendix \ref{sec:restrBayesProof} for a proof). Note that we adopt the notation $d_{\kappa}$ for a decision under the 0-$\kappa$ loss. Note, also, that a constraint on the type I error rate above that of the FD would imply the TI-RBD to differ with respect to the BD only if prior information is favoring the alternative hypothesis: When prior information supports the null hypothesis, the BD would imply a lower maximum type I error rate than the FD. However, the maximum frequentist risk $\underset{\theta \in \Theta}{\sup} \ R(\theta, d_{\kappa}) = \max \{\beta(\theta_0), \kappa (1- \beta(\theta_0)) \}$ would still grow above the one induced by the FD, and thus the RBD and TI-RBD would differ (see Figure \ref{fig:risk_ex} for a graphical example).
It is worth noting that an asymmetry in terms of the relative importance of type I and type II error rates is thus introduced in formulation (\ref{eq:restrictedBayes}). 

The constant $\eta$ has to be selected such that the Bayes decision with respect to $\nu$ gives a frequentist risk at $\theta_0$ equal to the pre-specified type I error rate $\beta(\theta_0)$. In this formulation $\eta$ has to be identified numerically. Moreover, it would not explicitly quantify the amount of type I error rate inflation.
However, it has to be noted that for families of distributions with monotone likelihood ratio a UMP test exists at each chosen type I error rate level \citep{lehm1986}. This suggests that type I error rate restricted Bayes decisions may be obtained by modifying the rejection region of the frequentist test in the `direction' suggested by the historical information. Notably, such an approach could also be adopted to solve (\ref{eq:restrBayes}), as exemplified by \citet{hodg1952}.
Here, we discuss and show the applicability of such an approach to a wider set of situations targeting explicitly (\ref{eq:restrictedBayes}), and to different choices of the prior distribution $\pi$. Moreover, we elicit a weight assigned to prior information that can be analytically related to the maximum allowed type I error rate inflation.

\section{Compromise decision}
\label{sec:proposedAppr}

To highlight the relationship between maximum type I error rate, cost of each test decision and incorporation of prior information, recall that the BD $d_{\kappa}^{\pi}$ under prior $\pi$ is to reject $H_0$ if $P^{\pi}(\theta \leq \theta_0|\mathbf{y}) < \tau$, where $\tau= \kappa/(1+\kappa)$. Assuming $\beta^{\pi}(\theta)$ to be increasing for $\theta\leq \theta_0$, denote $\beta^{\pi}(\theta_0)$ the maximum type I error rate of such procedure; this quantity is typically not analytically available and depends on both the prior density $\pi$ and the cost ratio $\kappa$. However, for data distributions with monotone likelihood ratio and non-degenerate prior distributions, posterior densities are stochastically ordered with respect to the data outcomes \citep{whit1979,milg1981}. This means in turn that the decision to keep or reject $H_0$ is a monotone function of the data (through the sufficient statistics), and therefore the test is UMP with type I error rate $\beta^{\pi}(\theta_0)$.  Note that such result has been proven for fixed and non-degenerate priors. When the prior is dynamically adapted according to the observed commensurability of current and historical information, the resulting test may not be UMP. An extreme situation where UMP property of the Bayesian test is lost under a dynamic prior choice is shown in \citet{kopp2019}. 
Note, also, that it is possible to identify a prior $\pi_0$ such that $P^{\pi_0}(\theta \leq \theta_0|\mathbf{y})$ is equal to the frequentist $p$-value, i.e., the probability (under the data distribution $f$) to obtain a data outcome as or more extreme than the observed $\mathbf{y}$ under the null hypothesis, in one-sided testing problems concerning location parameters of monotone likelihood ratio data families \citep[][]{case1987}. If such a prior $\pi_0$ is adopted for the analysis, then we have that the Bayes decision $d_{\kappa}^{\pi_0}$ induces a uniformly most powerful test with type I error rate $\beta^{\pi_0}(\theta_0)=\tau$, which exactly coincides with posterior probability threshold. 

The normal prior-normal likelihood is a commonly reported example which allows to show such parallelism between frequentist a Bayesian test decisions \citep[see e.g.][]{berg1985}. In particular, assume prior $\pi = N(\mu_{\pi}, \sigma_{\pi})$ and data $\bar{y} \sim N(\theta, \sigma/\sqrt{n})$, $H_0$ is rejected if $P^{\pi}(\theta \leq \theta_0|\mathbf{y}) < \tau$, i.e., if
\begin{equation}
\frac{\sqrt{n}}{\sigma} \left( \bar{y} - \theta_0 \right) >  \frac{\sigma (\theta_0 - \mu_{\pi})}{\sqrt{n} \sigma^2_{\pi}}  +  z_{1-\tau} \sqrt{1+\frac{\sigma^2}{n \sigma_{\pi}^2}},
\label{eq:bound}
\end{equation}
where $z_{1-\tau}$ denotes the $1-\tau$ quantile of the standard normal distribution. Denoting by $z^\pi$ the right-hand side of (\ref{eq:bound}), it follows that the type I error rate of the procedure equals $\beta^{\pi}(\theta_0)=1- \Phi(z^\pi)$, where $\Phi$ denotes the cumulative distribution function of the standard normal distribution. However, the same type I error rate as the one obtained under prior $\pi$ and posterior threshold, $\tau$, can be obtained in an analysis under a vague prior $\pi_0 = N(\mu_{\pi_0}, \sigma_{\pi_0})$, with $\sigma_{\pi_0} \rightarrow \infty$, by using a threshold $\tau^\pi = 1- \Phi(z^\pi)$ (this can be seen by replacing $\sigma_{\pi}$ with $\sigma_{\pi_0}$  and $z_{1-\tau}$ with $z_{1-\tau^\pi}$ in (\ref{eq:bound})). The  threshold $\tau^\pi$ coincides with the type I error rate of the test, and can be seen as induced by a revised cost ratio $\kappa^\pi$ which takes into account the prior probabilities assigned to each hypothesis by the informative prior distribution. It has therefore been suggested as a `rationally chosen' type I error rate for frequentist tests \citep{berg1985}. 

These observations suggest that a compromise decision may be reached either by tuning the frequentist test type I error rate level or, equivalently, the posterior probability threshold for rejection under $\pi_0$, so that it belongs to an interval which includes the standard type I error rate value $\beta^{\pi_0}(\theta_0)$, e.g. 0.025 or 0.05, and the prior-induced type I error rate value $\beta^{\pi}(\theta_0)$ as extremes.
We therefore achieve a `compromise decision' (CD) between the Bayes and frequentist decisions by defining a posterior decision threshold of the type 
\begin{equation}
\tau^w = (1-w) \tau + w \tau^\pi,
\label{eq:typeIlin}
\end{equation}
where $w \in [0,1]$. For an analysis undertaken under $\pi_0$, the type I error rate is thus equal to $\beta^w(\theta_0)=\tau^w$.

\subsection{Dynamic borrowing}

A major difficulty in most of the available borrowing approaches is the choice of the weight, or more generally, the degree of trust assigned to historical information. The compromise solution has the advantage that it relates such weight $w$ to the type I error rate inflation, but this would still require a prior assessment of the trust assigned to historical information. A reasonable solution would be to choose the weight based on \textit{a priori} considerations on the commensurability of the historical and current study populations as measured by, e.g., relevant covariates, 
but any other measure of discrepancy between the historical and current population may be used as well. 

Here we propose a dynamic approach in which the weight $w$ is defined according to the agreement between the informative and vague prior distribution analysis in supporting the alternative hypothesis. Formally,  
\begin{equation}
\hat{w} = 1-|P^{\pi}(\theta > \theta_0|\mathbf{y})-P^{\pi^*_{0}}(\theta > \theta_0|\mathbf{y})|,
\label{eq:weight}
\end{equation}
where $\pi^*_{0}$ is a prior specification with the same variance/informativeness as the informative prior, but with prior location agreeing with the observed data. Tuning the non-informative prior variance to match that of the informative prior one allows achieving full borrowing when no heterogeneity is observed between the informative prior and observed data location.
Note that $\hat{w}$ would only be affected by the similarity between the informative and modified vague prior analysis in the support given to the alternative hypothesis. This is in contrast with typical dynamic borrowing mechanisms, and is specifically tailored to testing: If interest lies in estimation, a different discounting mechanism should be adopted to avoid potentially large bias.  Note, also, that since the weight choice is in this case dependent on the current data outcome, the explicit relationship between the resulting weight and the type I error rate (\ref{eq:typeIlin}) cannot be applied and the exact type I error rate would have to be calculated via simulation. This can be seen by noting that the current data contribute to both sides of (\ref{eq:bound}) also after replacing $\sigma_{\pi}$ with $\sigma_{\pi_0}$, since $z_{1-\tau}$ is now replaced by the data-dependent $z_{1-\tau^{\hat{w}}}$. 
However, it is still possible to combine a dynamically estimated weight and a stricter control of the type I error rate, i.e., we choose 
\begin{equation*}
\tau^{\hat{w}} = \min[(1-\hat{w}) \tau + \hat{w} \tau^\pi, \tau^{bound}],
\label{eq:threshold}
\end{equation*}
where $\tau^{bound}$ is the maximum allowable type I error rate set by, e.g., regulatory requirements. This effectively constraints type I error rate also with a data-dependent weight: The actual type I error of the procedure can be lower, but it is guaranteed not to cross a set acceptability boundary. Indeed, if $z_{1-\tau^{\hat{w}}} \leq z_{1-\tau^{bound}}$, the type I error rate of the procedure will always be in $[0,\tau^{bound}]$. 

Finally, note that another situation in which (\ref{eq:typeIlin}) would not directly provide the type I error rate value of the procedure is one in which historical/external information is considered random (not yet realized or unblinded). This situation will be analytically investigated in \citeauthor{kopp2022} (in preparation).

\subsection{Connection with other robust borrowing approaches}

The CD $d_{\kappa^w}^{\pi_0}$ for any chosen fixed $w$ is also a TI-RBD satisfying (\ref{eq:restrictedBayes}) for an appropriately chosen $\eta$. To see this, note that both decisions are induced by a fixed non-degenerate prior distribution, and therefore they are a monotone function of the data. When $\eta=w=1$ prior information is fully incorporated, the same data threshold is induced by both decisions, and the test has type I error rate $\beta^{\pi}(\theta_0)$. For $w < 1$, data thresholds, and therefore test type I error rates, induced by the same values of $w$ and $\eta$ will however tend to differ, as exemplified in Section \ref{sec:simexample}.

An additional well-known approach reaching a compromise between FDs and BDs is represented by a BD with respect to a prior arising a mixture of the informative prior distribution $\pi$ and a vague or weakly informative prior distribution $\pi_{r}$ elicited with the aim of robustifying the analysis, i.e., $\gamma = (1-\epsilon) \pi_{r} + \epsilon \pi$, $\epsilon \in [0,1]$. Such a prior has been referred to as the `robust mixture prior' in the literature \citep[e.g.][]{schm2014}. Here we denote the decision induced by such prior $d_{\kappa}^{\gamma}$ as `robust mixture decision' (RMD). If $\pi_{r}$ is the point mass at $\theta_0$, then $\gamma=\nu$ for $\epsilon=\eta$. In all other cases, one cannot directly assume that the RMD will be equivalent to the TI-RBD for an appropriate choice of $\epsilon$. However, the robust mixture prior is a fixed and non-degenerate prior, thus, if the data outcome belongs to an exponential family distribution, we can again conclude that it induces UMP tests for any choice of $\pi_{r}$. Typically, however, as in the case of the CD and TI-RBD, the type I error rate of the test will differ for the RMD and TI-RBD for the same value of $\epsilon$ and $\eta$.

Several additional robust borrowing approaches have been proposed in the literature, e.g., power and commensurate prior formulations \citep{ibra2000,hobb2012}. When the adopted prior is fixed and non-degenerate, any such approach would lead to a UMP test at some (inflated, if the prior favours $H_1$) type I error rate level.

\subsection{Properties and sample size selection}

\paragraph{Sensitivity analyses}

Sensitivity analyses can be performed by exploiting the distinction between analysis and sampling prior \citep[see e.g.][and references therein]{cald2020}. The sampling prior represents the prior of the data-generating process, i.e., the prior under which parameters and consequently data samples are obtained.  For evaluation of frequentist operating characteristics, sampling priors are point-masses at different parameter values, typically $\theta_0$ as well as a $\theta$ value belonging to the alternative hypothesis support for power calculations. In the Bayesian context, sampling priors can belong to the same family as, e.g., the informative prior distribution. When the sampling and analysis prior coincide, the integrated risk is minimised. Sensitivity analyses are then carried out by adopting different sampling prior choices when computing the integrated risk, while test decisions are taken according to the analysis prior which will be used to fit the data. Formally, the integrated risk reads
 \begin{align}
r(\pi^s, d_{\kappa}(\mathbf{y}))=& \int_{-\infty}^{\theta_0} \beta(\theta) \pi^s(\theta) d\theta + \kappa \int_{\theta_0}^{\infty}  [1-\beta(\theta)] \pi^s(\theta) d\theta,
 \label{eq:intrisk}
 \end{align}
where $\pi^s$ denotes the sampling prior and $d_{\kappa}$ the test decision function under the 0-$\kappa$ loss induced by the analysis prior of choice. If $\pi^s=\pi$, the CD with weight $w$, $d^{\pi_0}_{\kappa^w}$, is suboptimal for (\ref{eq:intrisk}) unless $w=1$, but more robust in case the `true' sampling prior differs from $\pi$. To perform sensitivity analyses, the decision can therefore be fixed to the CD $d^{\pi_0}_{\kappa^w}$, while the sampling prior $\pi^s$ may be varied. 
As a smaller $w$ will result in solutions closer to the frequentist ones, the integrated risk would be less influenced by a discrepancy between the sampling prior and the prior obtained from historical information than if the BD $d^{\pi}_{\kappa}$ were adopted.

\paragraph{Integrated risk inflation}

We can also compute the `relative saving loss' ($RSL$), i.e., the ratio between the integrated risk inflation under the CD, and under the FD. \citet{efro1971} define it as
\begin{equation}
RSL (\pi) = \frac{r(\pi, d^{\pi_0}_{\kappa^w}(\mathbf{y}))-r(\pi, d^{\pi}_{\kappa}(\mathbf{y}))}{r(\pi, d^{\pi_0}_{\kappa}(\mathbf{y}))-r(\pi, d^{\pi}_{\kappa}(\mathbf{y}))},
\label{eq:RSL}
\end{equation}
where recall that $\pi_0$ represents the prior for which the posterior probability of the alternative is equal to the $p$-value. Note that in this formulation the risk is computed under $\pi^s=\pi$, and thus is minimised when fully incorporating prior information.
The $RSL$ is equal to 1 when all prior information is discarded, meaning that all advantages in terms of integrated risk minimisation (provided that $\pi^s=\pi$) are lost, while it is equal to 0 when all prior information is incorporated. The RSL focuses on the global integrated risk, i.e. it sums losses in both type I and type II error rate averaged with respect to the prior $\pi$. 

\paragraph{Sample size selection}

Sample size selection can be naturally incorporated in the proposed framework. We define expected power as power averaged with respect to the sampling distribution truncated at $\theta_0$, i.e., $\int_{\theta_0}^{\infty} \{1-\beta(\theta)\} \pi^s(\theta | \theta>\theta_0) d\theta$. One solution is thus to sample until expected power (either under the informative prior, or across a set of `realistic' sampling priors) has reached a desired level. 
As type I error rate is controlled throughout, this is of particular relevance in set-ups where type I error rate control is required by the regulators, while a sponsor is rather interested in maximising expected power.

\section{Example}
\label{sec:simexample}

In this section we illustrate the reviewed and proposed borrowing approaches focusing on a normal outcome. An analogous study for a binomial outcome is presented in the Supplementary Material.  We consider the set of hypotheses $H_0: \theta \leq \theta_0$ versus $H_1: \theta > \theta _0$. We assume $\kappa=0.025/0.975$, so that $\tau=\kappa/(1+\kappa)=0.025$, which corresponds in turn to the type I error rate under $\pi_0$. 

Let $\bar{y}$ denote the mean of independent and identically distributed observations from a $N(\theta,\sigma=1)$ distribution. Let $\pi$ denote the informative prior distribution, which is assumed to be $N(\mu_{\pi},\sigma_{\pi})$, and arising from $n_0=50$ historical observations with mean $\mu_{\pi}=0.25$ and standard deviation $\sigma$, i.e. $\sigma_{\pi}=1/\sqrt{50}$. Moreover, let $\theta_0=0$. As shown in Section \ref{sec:proposedAppr} the FDs are in this case induced by a normal prior $\pi_0$ with variance approaching infinity. Assuming $\pi_0$ to be a $N(0,100)$ provides in practice a good approximation for our purposes. Further, we explore the impact of decisions under two mixture prior specifications: The RMDs-Unit and RMDs-Vague. The RMDs-Unit are induced by $\gamma^{u} = (1-w) N(\mu_{\pi},1) + w N(\mu_{\pi},\sigma_{\pi})$, where the weakly informative component is taken to be a unit-information prior \citep{kass1995}, i.e., its variance is equal to that of a single observation. For the RMDs-vague, we consider decisions under $\gamma^{v} = (1-w) N(\mu_{\pi},100) + w N(\mu_{\pi},\sigma_{\pi})$ instead.
Let the TI-RBDs be induced by  $\nu= (1-w) \mu_0 + w N(\mu_{\pi},\sigma_{\pi})$, where $\mu_0$=$\delta_{\theta_0}$, and the CDs by $\pi_0$ with posterior probability threshold for rejection equal to $\tau^w=(1-w) \tau + w \tau^\pi$. Finally, in the CD-Adapt approach, the decision is undertaken under $\pi_0$, with threshold $\tau^{\hat{w}}=\min(0.15, (1-\hat{w}) \tau + \hat{w} \tau^\pi)$, with $\hat{w}$ as in (\ref{eq:weight}), and the constraint on $\tau^{\hat{w}}$ ensures that type I error rate is always below 0.15. Note that, to facilitate comparisons, the weight assigned to historical information is kept constant and equal to $w$ across all robust approaches with the exclusion of `CD-Adapt'. A summary of the implemented prior assumptions and decisions is provided in Table \ref{tab:approaches}.

\begin{table}[h!]
\setlength{\tabcolsep}{0.35em}
\centering 
 \begin{tabular}{|l|c|c|c|} 
 \hline
Decision & Prior & Threshold & Type I error rate \\
 \hline
 FD & $\pi_0=N(0,100)$ & $\tau=0.025$ & $\approx \tau$ \\
 BD &  $\pi=N(0.25,\frac{1}{\sqrt{50}})$ & $\tau=0.025$ & $\tau^\pi$ \\
 CD & $\pi_0=N(0,100)$ & $(1-w) \tau + w \tau^\pi$ & $\approx (1-w) \tau + w \tau^\pi$\\
 CD-Adapt & $\pi_0=N(0,100)$ & $(1-\hat{w}) \tau + \hat{w} \tau^\pi$ & to be eval. ($\leq 0.15$)\\
 RMD-Unit & $\gamma^{u} =(1-w) N(0.25,1) + w N(0.25,\frac{1}{\sqrt{50}})$ & $\tau=0.025$ & to be eval.\\
  RMD-Vague & $\gamma^{v} = (1-w) N(0.25,100) + w N(0.25,\frac{1}{\sqrt{50}})$ & $\tau=0.025$ & to be eval.\\
 TI-RBD & $\nu= (1-w) \mu_0 + w N(0.25,\frac{1}{\sqrt{50}})$ & $\tau=0.025$ & to be eval.\\
 \hline
 \end{tabular}
 \caption{List of implemented approaches.}
 \label{tab:approaches}
\end{table}

\begin{figure}[h!]
   \centering
   \includegraphics[width=1\textwidth]{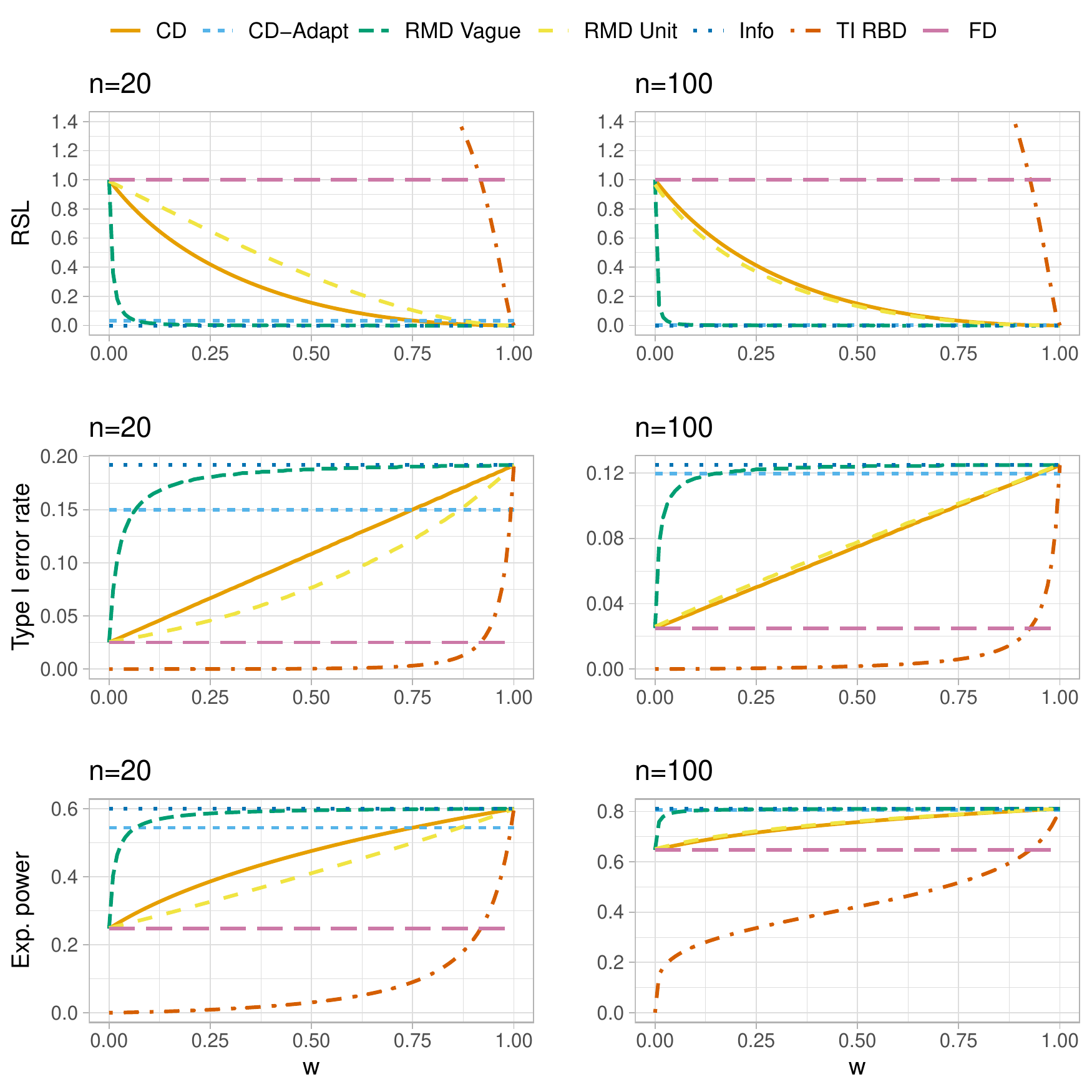}
   \caption{$RSL$, type I error rate and expected power for different test decisions, for varying weight assigned to historical information $w$, and for $n=20$ (left panels) and $n=100$ (right panels). The $RSL$ range is cut at 1.4 for plot optimisation; TI-RBD would extend well above the displayed range for small values of $w$.}
\label{fig:norm_err_f}
\end{figure}

Figure \ref{fig:norm_err_f} compares relative saving loss $RSL(\pi)$ in (\ref{eq:RSL}), type I error rate, and expected power of the different test decisions, i.e., the CD, the RMDs, the TI-RBD, CD-Adapt, and, as reference, the BD and the FD. The weight for CD, RMDs and TI-RBD is varied from 0, corresponding to no borrowing, to 1, corresponding to full borrowing. The results for CD-Adapt are averaged across data outcomes (and, thus, weights $\hat{w}$). We first notice that the TI-RBD can induce type I error rate and expected power levels well below those of the FD, even for relatively large weight assigned to the informative prior component. This, in turn, induces a sharp increase in expected power and decrease in $RSL$ only when $w$ is larger than approximately 0.9. Note the the TI-RBD is not in itself a proposed analysis strategy, and is just added for completeness following (\ref{eq:restrictedBayes}). The RMD-Unit and the CD show a strong similarity, particularly for $n=100$. Note that the RMD-Unit does not induce exactly the same results as the FD when $w=1$ due to the fact that the robust component is taken to be $N(\mu_{\pi},1)$. It is of interest to note, however, that the closeness of the CD and RMD-Unit depends quite heavily on the choice of the robust component. As can be observed in the graphs, under the RMD-Vague, historical information is effectively only discarded for very small values of $w$. This is a known phenomenon and related to the computation of the posterior weights: a very dispersed component induces a small marginal likelihood and thus causes the informative component to be favoured, irrespective of its commensurability to the data. For the same reason, vague prior components are not recommended when performing point-null hypothesis testing \citep{bartl1957,kass1995}. 

By construction, the CD the type I error rate of the CD varies linearly with the weight $w$. The RMD-Unit follows this pattern very closely in this example. An approximately linear relationship between the weight assigned to prior information and the change in the operating characteristic on which a compromise is sought seems a desirable feature which guarantees a meaningful interpretation of the weight itself. We believe that such `degree of linearity' can indeed be a useful measure to investigate for any borrowing mechanism, particularly when analytic relationships are not available.
The CD-Adapt approach ultimately leads to values identical to those of the  CD approach with $w= 0.75$ for $n=20$, and approximately $w=0.95$ for $n=100$. While this result is in general not guaranteed, it shows that in this case it ultimately avoids the need for a pre-specification of $w$ itself. Note that, for $n=20$, the results are driven by enforcement of the type I error bound of 0.15. 

\begin{figure}[h!]
   \centering
   \includegraphics[width=1\textwidth]{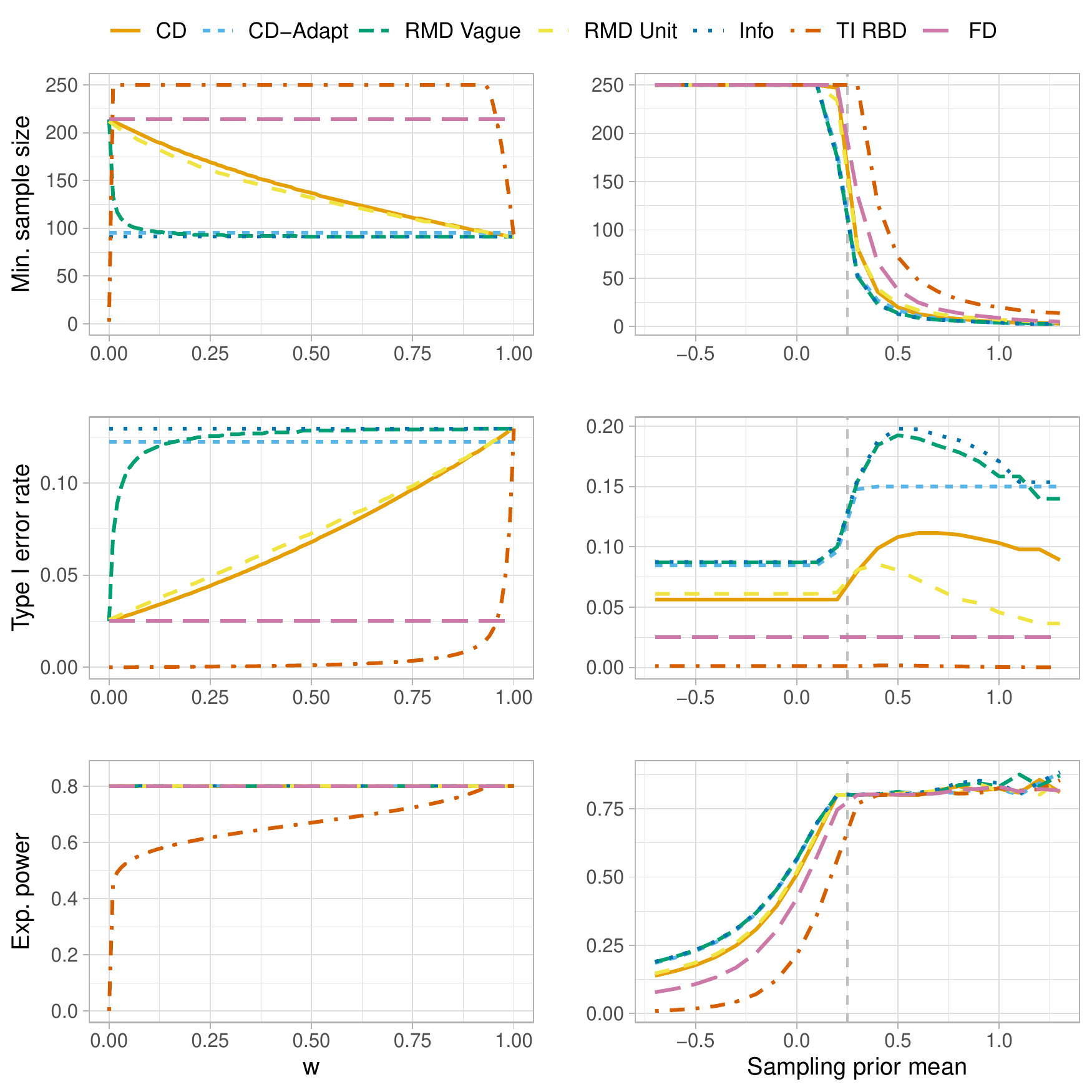}
   \caption{Minimum sample size to reach 80\% expected power, type I error rate and expected power for different test decisions, for varying weight assigned to historical information $w$, and fixed sampling prior $N(0.25,1/\sqrt{50})$ (left panels), and fixed $w=0.5$ and varying sampling prior mean (right panels). TI-RBD minimum sample size drop on the left panel is due to the fact that, for $w=0$, a point-mas prior at $\theta_0$ is obtained, therefore power would be equal to 0 for any sample size. The dashed grey vertical line shows the informative prior mean.}
\label{fig:norm_err_n}
\end{figure}

The left panels of Figure \ref{fig:norm_err_n} show how the required minimum sample size to achieve $80\%$ expected power would be affected by the choice of $w$ under the different borrowing approaches. Note that the maximum sample size has been truncated at 250. The minimum sample size decreases approximately linearly from the sample size required under no borrowing ($n=214$) to that required under full borrowing ($n=91$) under the RMD-Unit and the CD. Such a decrease mimics the increase in type I error rate, while expected power is generally controlled and above 0.8, as expected; the only exception is observed for the TI-RBD as sample sizes above 250 would be required under most $w$ values to achieve the expected power target. Again, the CD-Adapt approach corresponds to a weight of approximately 0.95 under CD.  Finally, RMD-Vague induces again a behaviour closer to the BD one for most weight choices.

To demonstrate how sensitivity analyses  in the context of sample size selection can be conducted, we fix $w=0.5$ for the TI-RBD, RMDs and CD approaches, and again a target expected power equal to 0.8 (note that expected power is computed with respect to the sampling prior truncated at $\theta_0$). We assume the sampling prior to be $N(\mu^{\pi^s},\sigma^{\pi^s})$, where $\sigma^{\pi^s}=\sigma_\pi$, and varying sampling prior mean. The top right panel of Figure \ref{fig:norm_err_n}  shows how the minimum required sample size is affected by the location of the sampling prior. Operating characteristics under the BD and FD are superimposed for comparison. Recall that the FD always controls type I error rate at 0.025, while the BD type I error rate varies according to the sample size, which is in turn determined by the expected power target. Recall also the we truncate the sample size at 250, so expected power would be below the target 0.8 when a larger sample size would be needed to achieve it.  We observe that historical information is beneficial in terms of expected power and thus, in turn, sample size savings can be achieved over the whole range of sampling prior means when prior do incorporate such information; however, a decrease in required sample size can induce a significant inflation in type I error rate. 
The CD behaves again closely to the RMD-Unit in terms of expected power and sample size requirements, but its   type I error rate is fully determined by the choice of $w$, and implies that, under any sample size and data generating mechanism, the inflation in type I error rate would be 50\% of the increase in type I error rate induced by the BD (for the same sample size). Finally, the CD-Adapt approach achieves sample size gains close to that of the informative and RMD-Vague prior distribution while maintaining the type I error rate, as planned, below 0.15.

\begin{figure}[h!]
   \centering
   \includegraphics[width=1\textwidth]{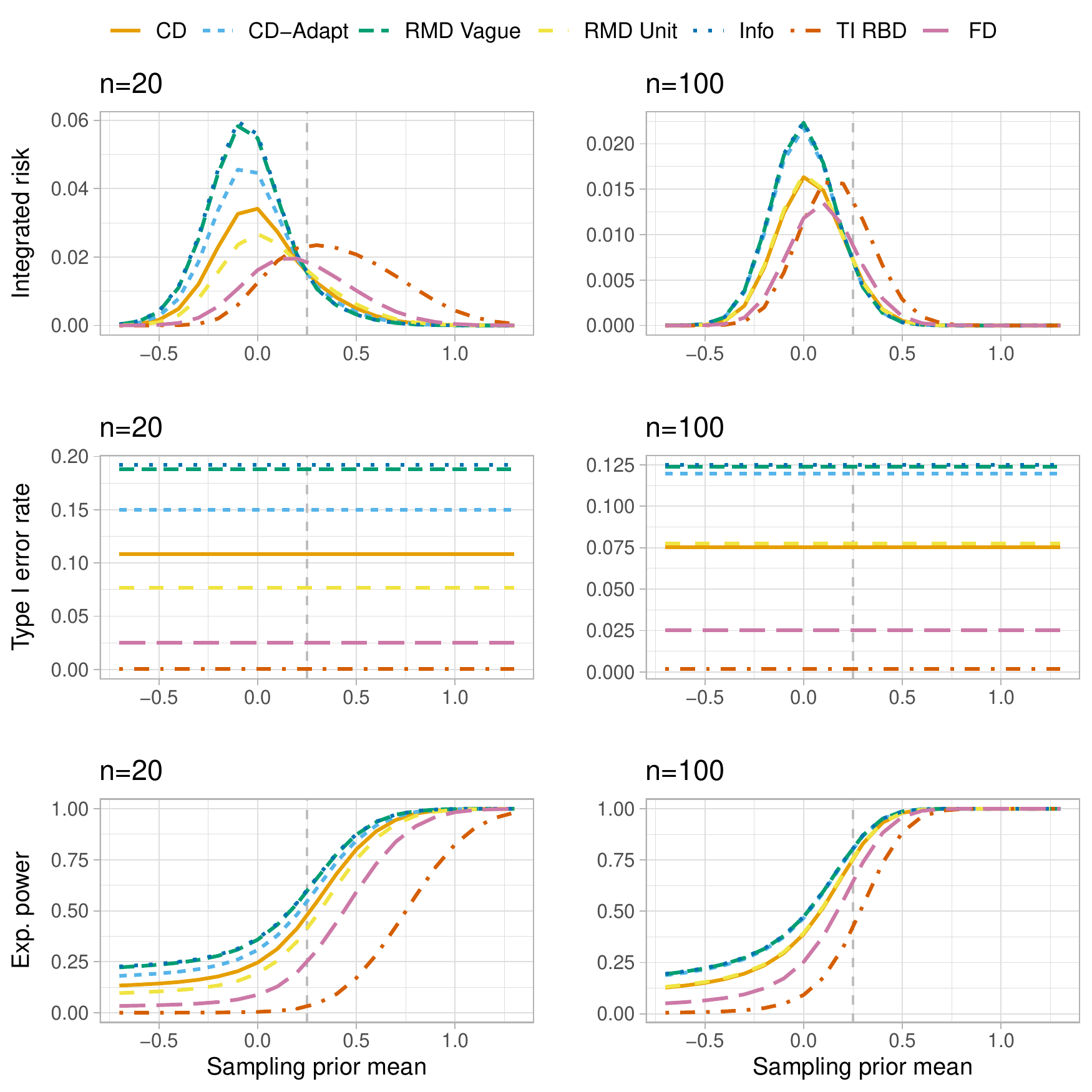}
   \caption{Integrated risk, type I error rate and expected power for different test decisions, for fixed $w=0.5$, varying sampling prior mean, and $n=20$ (left panels) and $n=100$ (right panels). The dashed grey vertical line shows the informative prior mean.}
\label{fig:norm_sens}
\end{figure}

Sensitivity analyses for the integrated risk, for sample size equal to 20 and 100, are shown in Figure \ref{fig:norm_sens}. For both sample sizes, the informative prior and RMD-Vague again achieve very close results, with the integrated risk being minimised when the sampling prior mean coincides with the informative prior mean, but with potential significant inflation otherwise. The FD has the lowest maximum integrated risk, although it loses some integrated risk gains as compared to decisions incorporating historical information, when the sampling and informative prior are consistent, as expected. 

\section{Conclusions}

In this work we have approached the problem of building a principled frequentist-Bayesian compromise decision to testing under a 0-$\kappa$ loss for location parameters of exponential family distributions in one-arm studies. The CD has a straightforward interpretability in that the role of historical information in the type I error rate inflation is made explicit through the modification of the cost ratio between type I and type II errors. Note that the approach does not require a Bayesian analysis strategy in itself. However, it does require a certain amount of trust in historical information. Under a fixed borrowing mechanism, the CD has been shown to be optimal in terms of minimisation of the integrated risk, subjected to type I error rate constraint.
If the Bayesian paradigm is fully embraced, no such compromise is necessary. However, full commensurability is a strong assumption and sensitivity analyses are often necessary to evaluate benefits and losses associated with borrowing of historical information under various heterogeneity scenarios \citep{viele2014}. In this context, the proposed approach provides a tool to perform such evaluations and cap type I error rate inflation at a pre-specified level. 

In analogy with several borrowing approaches, the amount of type I error rate inflation is tuned by the choice of a specific parameter, in our case the weight $w$, which represents the allowed proportion of type I error rate inflation, as compared to an analysis fully incorporating historical information. We have outlined possibilities for the choice of such a weight, and proposed a novel adaptive approach for its elicitation which explicitly focuses on testing.
We have compared the operating characteristics obtained when assigning the same weight to historical information in the (static) CD and the robust mixture prior approaches. Interestingly, similarity between the two approaches heavily depends on the choice of the robustifying component in the mixture prior, with, e.g., a unit-information one inducing much stronger similarities (thus, linearity in type I error rate inflation), than a very diffuse prior in the normal outcome case. 
More generally, a high degree of linearity between the weight assigned to historical information and the operating characteristic on which compromise is sought seems to us a desirable property of any borrowing mechanism, as it implies a good interpretability of the weight itself. We thus believe that such an assessment could be of broader interest and a potential topic for further research. 

We have provided tools for sensitivity analyses and sample size selection. We have primarily focused on reaching a target expected power, defined as average power with respect to the sampling distribution truncated and $\theta_0$. This is not the only option. The lower truncation boundary $\theta_0$ can be replaced, e.g., by a relevance threshold, or by the whole parameter range \citep[see][for a comprehensive review of measures in this context]{kunz2020}. If a decision-theoretic approach is fully embraced, a cost can be assigned to each sample and sample size can be added and optimised via minimisation of the integrated risk itself. Note, however, that sampling until a certain target is reached is not `cost-free'; rather, implicit costs are assigned \citep{lind1997,cald2020}. 

We have focused on one-arm studies. Extension to two-arm situations poses no difficulties if a single prior is elicited on the difference between the treatment and control mean: When Normal outcomes are considered, reduction to a one-arm design is straightforward and the method can be directly applied. A situation requiring further study is when separate priors are elicited for the treatment and the control arm, and will be the focus of future research.

\vspace{1cm}
\noindent {\bf{Conflict of Interest}}

\noindent {\it{The authors have declared no conflict of interest.}}

\section{Appendix}

\subsection*{A.1.\enspace Restricted Bayes solution with constraint on maximum type I error}
\label{sec:restrBayesProof}

The following theorem and proof follows \citet{hodg1952}, with minor adaptations.

Theorem. Let $\nu= \eta \pi + (1-\eta) \mu_0$ and $d^{\nu}$ the Bayes solution with respect to $\nu$. If $d^{\nu}$ is such that $\int_\Theta R(\theta, d^{\nu}) \mu_0(\theta) d\theta = R(\theta_0,d^{\nu})$, then it also minimizes 
\begin{equation*}
    \eta \int_\Theta R(\theta, d) \ \pi(\theta) \ d\theta + (1-\eta) R(\theta_0, d).
\end{equation*}

Proof. Let $d$ be any decision, then 
\begin{align*}
    &\eta \int_\Theta R(\theta, d) \ \pi(\theta) \ d\theta + (1-\eta) R(\theta_0, d) \geq \eta \int_\Theta R(\theta, d) \ \pi(\theta) \ d\theta + (1-\eta) \int_\Theta R(\theta, d) \mu_0(\theta) d\theta \\
    \geq &\eta \int_\Theta R(\theta, d^{\nu}) \ \pi(\theta) \ d\theta + (1-\eta) \int_\Theta R(\theta, d^{\nu}) \mu_0(\theta) d\theta =  \eta \int_\Theta R(\theta, d^{\nu}) \ \pi(\theta) \ d\theta + (1-\eta) R(\theta_0, d^{\nu}).
\end{align*}

It also follows that $\mu_0$ is the point-mass density at $\theta_0$.

\bibliographystyle{apalike}
\bibliography{refs}

\end{document}


\begin{center}
\huge{\textbf{Supplementary material}}\\

\vspace{1cm}
\Large
\textit{Robust incorporation of historical information with known type I error rate inflation}\\
\centering Silvia Calderazzo \& Annette Kopp-Schneider
\end{center}
\vspace{1cm}

\begin{figure}[h!]
   \centering
   \includegraphics[width=1\textwidth]{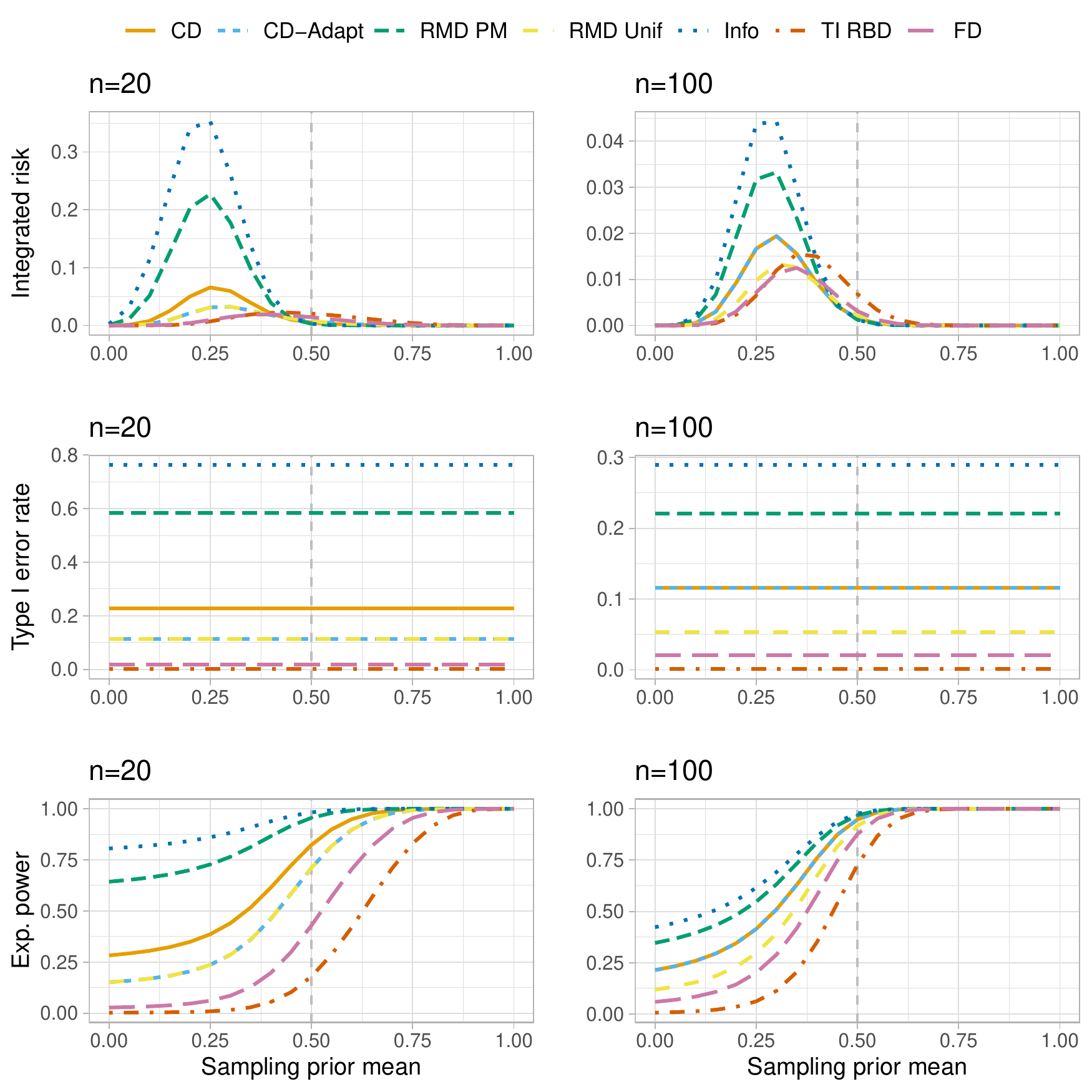}
   \caption{Integrated risk, type I error rate and expected power for different test decisions, for fixed $w=0.5$, varying sampling prior mean, and $n=20$ (left panels) and $n=100$ (right panels). The dashed grey vertical line shows the informative prior mean.}
\label{fig:binom_sens}
\end{figure}

\normalsize{
\section*{Binomial outcome}
\label{sec:binom}

Here we assume that $y$ is an observation from a $Bin(n,\theta)$ random variable. The informative prior $\pi$ is taken to be a conjugate distribution $Beta(1+a_{hist},1+b_{hist})$, where $a_{hist}=20$ and $b_{hist}=20$ represent the number of historical successes and failures, respectively, therefore $n_0=40$. For the one-sided testing problem considered, i.e.,  $H_0: \theta \leq \theta_0$ versus $H_1:\theta> \theta_0$, a $Beta(0,1)$ implies equivalence of posterior probabilities of the null hypothesis and p-values \citep{leco2007,kopp2019b}. As for normal outcomes, we use a close approximation, i.e., $\pi_0$ is taken to be a $Beta(0.001,1)$. Moreover, let $\theta_0=0.3$. We assume the RMD-unif to be induced by a prior $\gamma^{unif}= (1-w) Beta(1,1) + w Beta(1+a_{hist},1+b_{hist})$, the RMD-PM by a prior $\gamma^{PM}= (1-w) Beta(0.001,1) + w Beta(1+a_{hist},1+b_{hist})$, and the TI-RBD to arise under a distribution of the type $\nu= (1-w) \mu_0 + w Beta(1+a_{hist},1+b_{hist})$. The CD is obtained under $\pi_0$ with posterior probability threshold for rejection equal to $\tau^w=(1-w) \tau + w \tau^\pi$. In the CD-Adapt case, we take $\hat{w}=1-|P^{\pi}(\theta > \theta_0|\mathbf{y})-P^{\pi^*_{0}}(\theta > \theta_0|\mathbf{y})|$, where $\pi^*=Beta(1+y(1+n_0/n),1+n+n_0 - y(1+n_0/n))$.

Figure \ref{fig:binom_err_f} shows a comparison analogous to the normal case for the $RSL$, the maximum type I error rate, and expected power. The TI-RBD once again crosses the FD-induced operating characteristics at a relatively high weight $w$. The RMD-unif and the CD induce again similar results, although we observe in this case a more marked tendency of the RMD-unif to achieve smaller type I error rates (and thus lower expected power and higher $RSL$) than the CD for a given weight. Instead, the RMD-PM follows more closely the BD behaviour, moving far away from the (approximate due to discreteness) linear relationship between the weight and the maximum type I error rate of the CD. As discussed in the normal outcome example, this suggests a poor interpretability of the weight and thus a less appealing borrowing approach. The CD-Adapt approach is additionally super-imposed for comparison. When $n=20$, it corresponds to the CD approach with weight around 0.25, while it corresponds to a weight around 0.4 for $n=100$. As no true underlying conflict is present in this first set of situations, both results are driven by the upper boundary to type I error rate set at 0.15 (and sampling variability).

Similar considerations on the comparison between the RMD-Unif and the CD can be applied to the sample size requirement to achieve $80\%$ expected power, shown in the left panels of Figure \ref{fig:binom_err_n}. Note that the sample size requirement varies between $n=1$ for the BD, which however induces a type I error rate equal to 1, and $n=71$ when all prior information is discarded. The TI RBD would reach much higher sample size requirements for most $w$ values due to its point-mass component. The CD-Adapt reaches a maximum type I error rate of 0.11 and requires 34 samples. Sensitivity analyses for varying sampling prior $Beta(1+a_{s},1+b_{s})$ under the constraint $a_{s} + b_{s}=40$, and fixed $w=0.5$ for CD, RMDs and TI RBD, are shown in the right panels. Consistently with the fixed sampling prior results at $w=0.5$, the sample size requirement of the BD is uniformly lower, and the TI-RBD one uniformly higher, than all other approaches. The RMD-PM follows very closely the BD behaviour. The RMD-unif, on the other hand, leads to a higher sample size requirement than the CD but at the same time lower type I error rate inflation. The CD-Adapt approach has  sample size requirements closer to that of the BD and RMD-PM for small sampling prior means while it approaches the RMD-Unif as the alternative hypothesis becomes more likely under the sampling prior distribution.

Sensitivity analyses for the integrated risk, for sample size equal to 20 and 100, are shown in Figure \ref{fig:binom_sens}. The BD achieves the smallest integrated risk among all decisions when it coincides with the sampling prior. On the other had, its integrated risk also achieves the overall maximum across the sampling prior mean range. The FD is has the lowest maximum integrated risk, although it loses some integrated risk gains when the sampling and informative prior are consistent, as expected.

\begin{figure}[h!]
   \centering
   \includegraphics[width=1\textwidth]{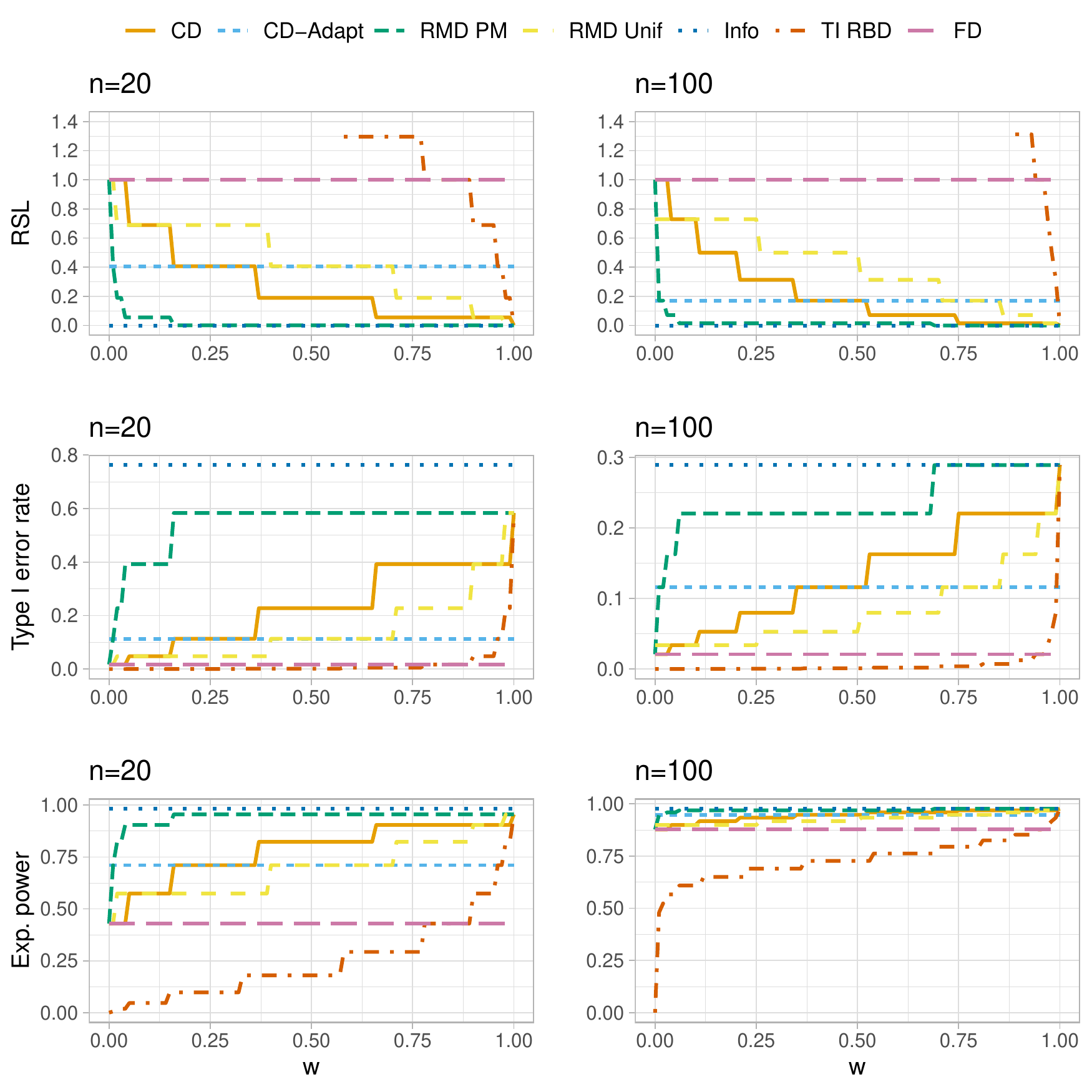}
   \caption{$RSL$, maximum type I error rate and expected power for different test decisions, for varying weight assigned to historical information $w$, and $n=20$ (left panels) and $n=100$ (right panels). The $RSL$ range is cut at 1.4 for plot optimisation; TI-RBD would extend well above the displayed range for small values of $w$.}
\label{fig:binom_err_f}
\end{figure}

\begin{figure}[h!]
   \centering
   \includegraphics[width=1\textwidth]{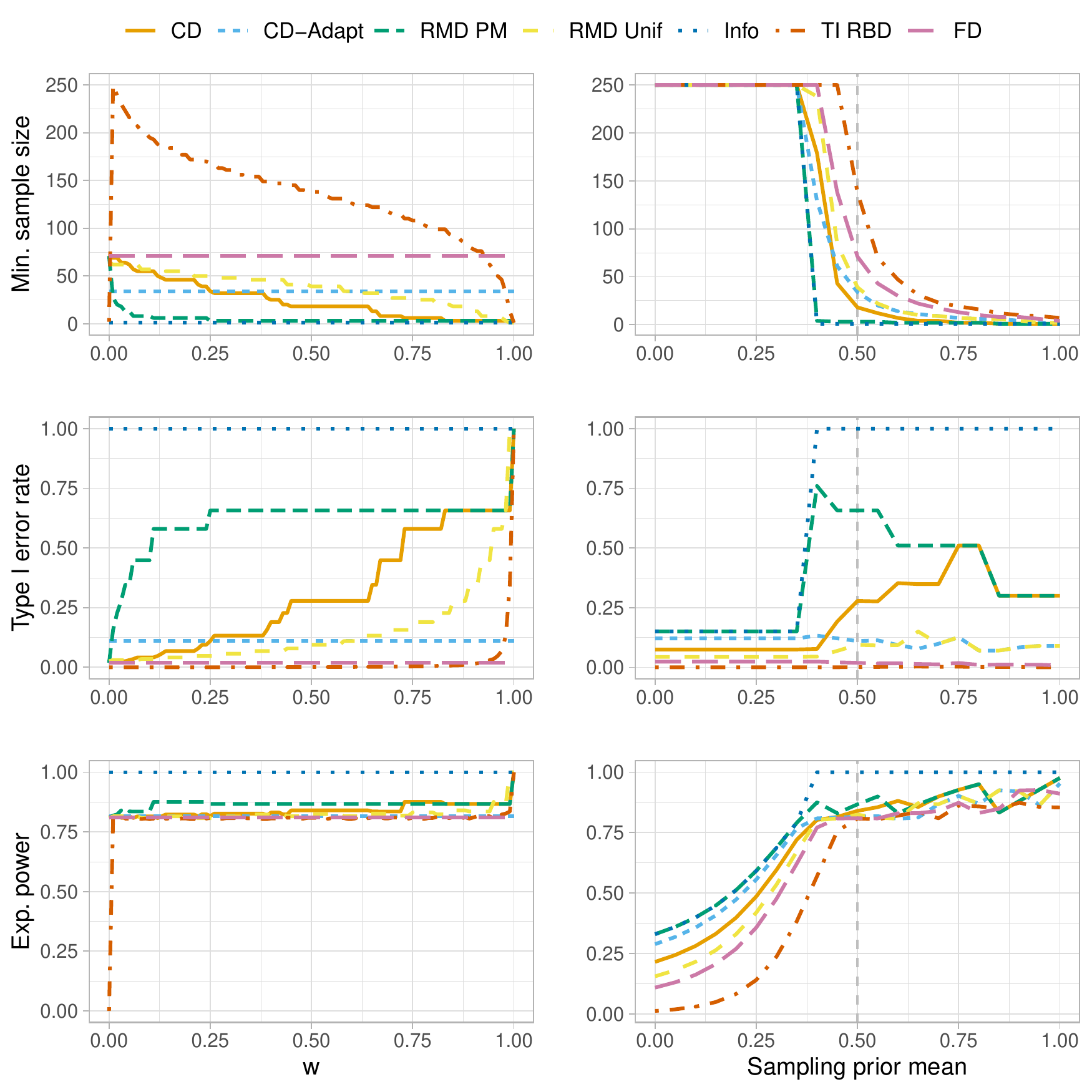}
   \caption{Minimum sample size to achieve 80\% expected power,  maximum type I error rate and expected power for different test decisions for varying weight assigned to historical information $w$ and fixed sampling prior $Beta(21,21)$ (left panels), and fixed $w=0.5$ and varying sampling prior $Beta(a_{s}+1,ab_{s}+1)$ under the constraint $a_{s} + b_{s}=40$ (right panels). TI-RBD minimum sample size drop on the left panel is due to the fact that, for $w=0$, a point-mas prior at $\theta_0$ is obtained, therefore power would be equal to 0 for any sample size.}
\label{fig:binom_err_n}
\end{figure}
}

\bibliography{refs}
\bibliographystyle{apalike}